\documentclass[conference]{IEEEtran}
\IEEEoverridecommandlockouts

\usepackage{cite}
\usepackage{amsmath,amssymb,amsfonts}
\usepackage{algorithmic}
\usepackage{graphicx}
\usepackage{textcomp}
\usepackage{xcolor}
\usepackage{svg}
\usepackage{amsmath}
\usepackage{subcaption}
\usepackage{mwe}
\usepackage[a4paper,
            left=0.75in,
            right=0.75in,
            top=0.7in,
            bottom=1in]{geometry}

\usepackage[font=small]{caption}

\def\BibTeX{{\rm B\kern-.05em{\sc i\kern-.025em b}\kern-.08em
    T\kern-.1667em\lower.7ex\hbox{E}\kern-.125emX}}
\begin{document}

\title{\vspace{-4mm} Generative AI-enabled Blockage Prediction for  Robust Dual-Band mmWave Communication\\ \vspace{-4mm}}

\author{\IEEEauthorblockN{Mohammad Ghassemi\textsuperscript{1}, Han Zhang\textsuperscript{1}, Ali Afana\textsuperscript{2}, Akram Bin Sediq\textsuperscript{2}, \\ and Melike Erol-Kantarci\textsuperscript{1}, \textit{Senior Member}, \textit{IEEE}}
\IEEEauthorblockA{\textit{\textsuperscript{1}School of Electrical Engineering and Computer Science, University of Ottawa, Ottawa, Canada} \\
\textit{\textsuperscript{2}Ericsson Inc., Ottawa, Canada}\\
Emails:\{mghas017, hzhan363, melike.erolkantarci\}@uottawa.ca,
\{ali.afana, akram.bin.sediq\}@ericsson.com}

\vspace{-10mm}}

\maketitle

\begin{abstract}
In mmWave wireless networks, signal blockages present a significant challenge due to the susceptibility to environmental moving obstructions. 
Recently, the availability of visual data has been leveraged to enhance blockage prediction accuracy in mmWave networks. In this work, we propose a Vision Transformer (ViT)-based approach for visual-aided blockage prediction that intelligently switches between mmWave and Sub-6 GHz frequencies to maximize network throughput and maintain reliable connectivity.
Given the computational demands of processing visual data, we implement our solution within a hierarchical fog-cloud computing architecture, where fog nodes collaborate with cloud servers to efficiently manage computational tasks.
This structure incorporates a generative AI-based compression technique that significantly reduces the volume of visual data transmitted between fog nodes and cloud centers.
Our proposed method is tested with the real-world DeepSense 6G dataset, and according to the simulation results, it achieves a blockage prediction accuracy of 92.78\% while reducing bandwidth usage by 70.31\%.
\end{abstract}

\section{Introduction}
The rapid development of wireless technology and the diversity of applications  have increased the need for higher data transmission rates. In 5G and 6G, using millimeter-wave (mmWave) frequency for communication has been found to be an efficient way to provide high data rates in line-of-sight (LOS) links \cite{rappaport2019wireless}. However, the mmWave communication is significantly impacted by blockages due to high penetration loss, leading to link degradation or disconnections in non-LOS (NLOS) conditions \cite{charan2022computer}. 
A solution to this problem can be proactively predicting blockages and switching from mmWave to sub-6 GHz frequency if a blockage is detected given that two bands are co-deployed. In such scenarios, accurate blockage prediction becomes key issue to ensure mmWave communication reliability \cite{elsayed2020radio}\cite{semiari2019integrated}.

In recent years, visual data have been applied to enhance the performance of the blockage prediction task for next-generation base stations (BSs) that are anticipated to have sensing capabilities
\cite{xie2023communication}\cite{khan2024semantic}\cite{farzanullah2024generative}. Vision Transformer (ViT) is a transformer-like model that is widely used to handle vision processing tasks \cite{dosovitskiy2020image}. By segmenting images into patches and processing them as sequences, ViT effectively extracts global context and intricate spatial relationships across an image \cite{gharsallah2024vit}.

The integration of visual data, typically characterized by large storage needs, imposes a considerable storage demand on the BS. Additionally, running ViT directly on the BS requires substantial computational resources, creating a processing burden that may not be feasible for real-time applications \cite{zhou2014optimized}. Therefore, we use a hierarchical fog-cloud computing system that performs initial processing tasks at the BS while leveraging the cloud server to handle more complex computations. These systems capitalize on the advantages of mobile edge computing (MEC) and mobile cloud computing (MCC) to manage computation-intensive applications \cite{cheng2016computation}\cite{nguyen2019joint}.
However, fog-cloud computing systems can experience significant overhead and impose a heavy signaling load, which may lead to quality of service (QoS) degradation \cite{liu2018cooperative}. 
In our proposed blockage prediction task, we utilize visual data, which constitutes a large portion of network traffic when transmitted directly to the cloud server for processing \cite{cisco2017}\cite{yang2017multimedia}. To address signaling challenges, a promising solution is to implement image compression techniques that can significantly reduce the size of transmitted data. Recent advancements in deep learning-based image codecs have outperformed traditional methods, enabling efficient compression of visual data while preserving quality. By leveraging these modern codecs, we can enhance data transmission efficiency \cite{yang2024lossy}.
These advanced codecs enhance data transmission efficiency by minimizing visual data and can be integrated into our framework \cite{yang2024lossy}.

To the best of our knowledge, no prior research has explored the integration of image compression with ViT-based blockage prediction for dual-band communication. The main contributions of this work are as follows: 

\begin{itemize}
\item We perform vision-aided blockage prediction to enable the BS to switch intelligently between mmWave and Sub-6 GHz bands. This approach significantly enhances the quality of communication by ensuring seamless connectivity in the presence of blockages.
\item To reduce the overhead of the fog-cloud system, we perform image compression and employ a generative AI encoding and decoding technique, as described in \cite{yang2024lossy}, to extract latent features from images collected at the BS. This compression reduces the data size, leading to decreased bandwidth usage on the transport link between the BS and cloud centers within a fog-cloud system. 
\end{itemize}

According to the simulation results, our proposed method decreases bandwidth usage by 70.31\%. For blockage prediction, our method achieves an accuracy of 92.78\%, outperforming the baseline transformer model, which achieves 86.23\% accuracy when using compressed images.

The structure of the paper is organized as follows: Section II presents the related work. Section III outlines the system model and problem formulation. Section IV details the blockage prediction framework. Section V presents simulation results and performance comparisons. Finally, Section VI concludes the study by summarizing the findings and implications of the proposed approach.

\section{Related work}
In recent years, there has been existing studies that discuss dual-band switching facilitated by blockage identification. For instance, \cite{semiari2019integrated} explores the integration of mmWave and Sub-6 GHz frequencies, emphasizing a flexible architecture that enables seamless switching between the two bands. Specifically, when a mmWave link encounters blockage or misalignment, the system reroutes data through the Sub-6 GHz band, ensuring continuous service and robust performance. These works highlight the importance of accurate blockage detection to enhance the stability of mmWave communications.
In addition, there has been an increasing focus on addressing link blockage issues, particularly in high-frequency wireless networks. Several studies have explored this area. 
In \cite{alrabeiah2020millimeter}, the authors proposed a machine learning-based method for proactively predicting mmWave link blockages using visual data from RGB cameras, utilizing the real-world DeepSense 6G dataset. The paper \cite{khan2024semantic} introduced a framework that employs computer vision techniques to extract semantic information from images, coupled with federated learning for distributed on-device learning to enhance blockage prediction in next-generation wireless networks. 
The authors in \cite{gharsallah2024vit} presented a novel ViT architecture for predicting blockages in 6G vehicular networks, leveraging multi-view information from connected vehicles to improve wireless communication reliability. \cite{alrabeiah2020deep} discussed a deep learning approach for predicting mmWave beams and blockages using sub-6 GHz channels.
However, these approaches can lead to high computational demands, making them impractical to implement directly on the BS in real-world environments. To address this challenge, a hierarchical fog-cloud computing system can be used. Fog-cloud computing has been successfully applied in other network applications, such as healthcare monitoring and emergency response systems, to reduce latency and manage data effectively at the network edge before offloading to the cloud \cite{dastjerdi2016fog}. Given the visual-based blockage detection scenario, the fog-cloud computing approach can similarly optimize network performance. 
This approach distributes the processing load by handling initial data processing at the fog nodes and offloading complex computations to the cloud, reducing the BS's computational burden while ensuring seamless communication through dynamic switching between mmWave and Sub-6 GHz frequencies based on blockage predictions.

On the other hand, image compression has been explored through some existing research. For instance, \cite{mentzer2020high} effectively learns a compact latent representation of images and reconstructs them with minimal visual loss. 
The authors in \cite{yang2024lossy} introduced a novel image compression method utilizing diffusion models, which maps images to a contextual latent variable while preserving essential features. 
While we adopt their approach for encoding images and reducing data transmission size, our work extends this by integrating the compressed visual data into blockage prediction and dual-band communication system.

\section{System Model and Problem Formulation}
This section outlines the dual-band communication model, formulates the blockage prediction problem, and details the deployment within a fog-cloud system, including encoding and decoding formulas to minimize overhead.
\begin{figure}
\setlength{\belowcaptionskip}{-12pt}
\centering
  \centering
  \includegraphics[width=0.8\linewidth]{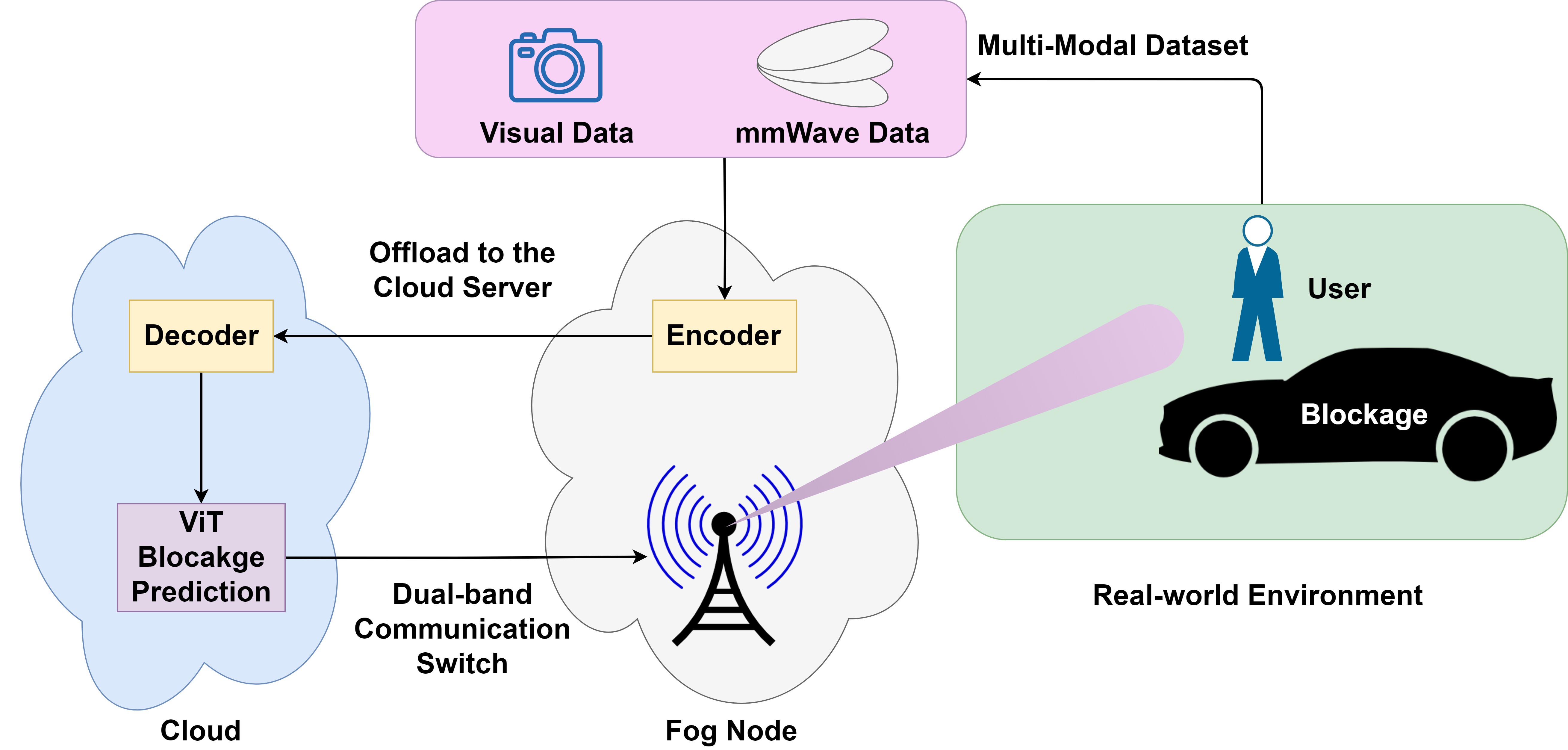}
  \captionof{figure}{System architecture for blockage prediction using visual-aided data.}
  \label{fig:image1}
\end{figure}

\subsection{Dual-Band Communication Model}
This study explores a system where a BS connects to a user in a busy environment with moving objects that may cause blockages \figurename{\ref{fig:image1}}. The BS is capable of utilizing both mmWave and Sub-6 GHz frequencies to establish a connection with the user. The dual-band channel model $h_k$ at time $t$ can be expressed as \cite{charan2022computer}:
\begin{equation}
    h_k[t] = (1 - a[t]) h_k^{\text{LOS}}[t] + a[t]h_k^{\text{NLOS}}[t],
\end{equation}
where $h_k^{\text{LOS}}$ and $h_k^{\text{NLOS}}$ denote the LOS and NLOS channel components, respectively. The binary variable $a[t] \in \{0, 1\}$ indicates the condition of the link at time $t$, where $a[t] = 1$ reflects an NLOS link, and $a[t] = 0$ represents a LOS link.

\subsection{System Model Problem Formulation} \label{subsectionSystemModel}
The objective is to predict whether the user will encounter a blockage within a future window of $r^{'}$ time instances based on data observed at the BS. Let $\mathbf{X}[t] \in \mathbb{R}^{W \times H \times C}$ be an image, where $W$, $H$, and $C$ represent width, height, and channels. Additionally, $\mathbf{p}[t] \in \mathbb{R}^{64}$ denotes the received mmWave power vector at time $t$.
The BS processes a sequence of samples, $\mathcal{V}[t]$, and it can be defined as $\mathcal{V}[\tau] = \{\mathbf{X}[t], \mathbf{p}[t]\}_{t = \tau - r + 1}^{\tau}$,
where $r \in \mathbb{N}$ represents the length of the input sequence to predict future link blockages. The goal is to forecast the blockage status $s[t]$ within a future window $r^{'}$ time instances, expressed as:
\begin{equation}
s[\tau] = 
\begin{cases}
1, & a[t] = 1, \; t \in \{\tau + 1, \ldots, \tau + r^{'}\} \\
0, & \text{otherwise}
\end{cases}
\end{equation}
where $0$ signifies that the user remains in a LOS state, and $1$ indicates a blockage within the upcoming $r^{'}$ instances.

To predict the future blockage status, we define function $f_{\Theta}$ that processes the observed sequence of $\hat{\mathcal{V}}[\tau]$, which will be defined in \ref{subsectionSystemModelC}, and generates a prediction of the future blockage status, $\hat{s}[\tau]$ as $f_{\Theta} : \hat{\mathcal{V}}[\tau] \rightarrow \hat{s}[\tau]$.

In this study, we employ a ViT to train the prediction function $f_{\Theta}$, optimizing the function parameters $\Theta$ from a dataset of labeled sequences $\mathcal{D} = \{(\hat{\mathcal{S}}_v, s_v)\}_{v=1}^{V}$. The optimization goal is to ensure high prediction accuracy across samples, which can be formulated as:
\begin{equation} \label{eq:4}
f_{\Theta^{*}}^{*} = \arg\max_{f_{\Theta}(\cdot)} \prod_{v=1}^{V} \mathbb{P}(\hat{s}_v = s_v|\hat{\mathcal{V}}_v).
\end{equation}

\subsection{Deployment in Fog-Cloud Infrastructure}\label{subsectionSystemModelC}
The system integrates a hierarchical fog-cloud infrastructure where visual and received mmWave power vector data are processed at the BS. The BS is equipped with sensors to monitor the surroundings, enabling it to collect sensing data that can be used to predict potential link blockages caused by moving objects. Each image is processed by an encoder at the fog node, producing a compressed representation $\mathbf{z}[t] \in \mathbb{R}^{C_H}$. The compressed data is transmitted to the cloud server, where a decoder reconstructs the received sequence of images $\mathbf{Y}[t]$. 

Therefore, the received data before ViT processing is $\hat{\mathcal{V}}[\tau] = \{\mathbf{Y}[t], \mathbf{p}[t]\}_{t = \tau - r + 1}^{\tau}$.
For this analysis, we assume the communication link between the fog node and the cloud center is noise-free.

\section{Communication-Efficient Blockage Prediction\label{section4}}
\begin{figure*}
\setlength{\belowcaptionskip}{-12pt}
\centering
  \centering
  \includegraphics[width=1.0\linewidth]{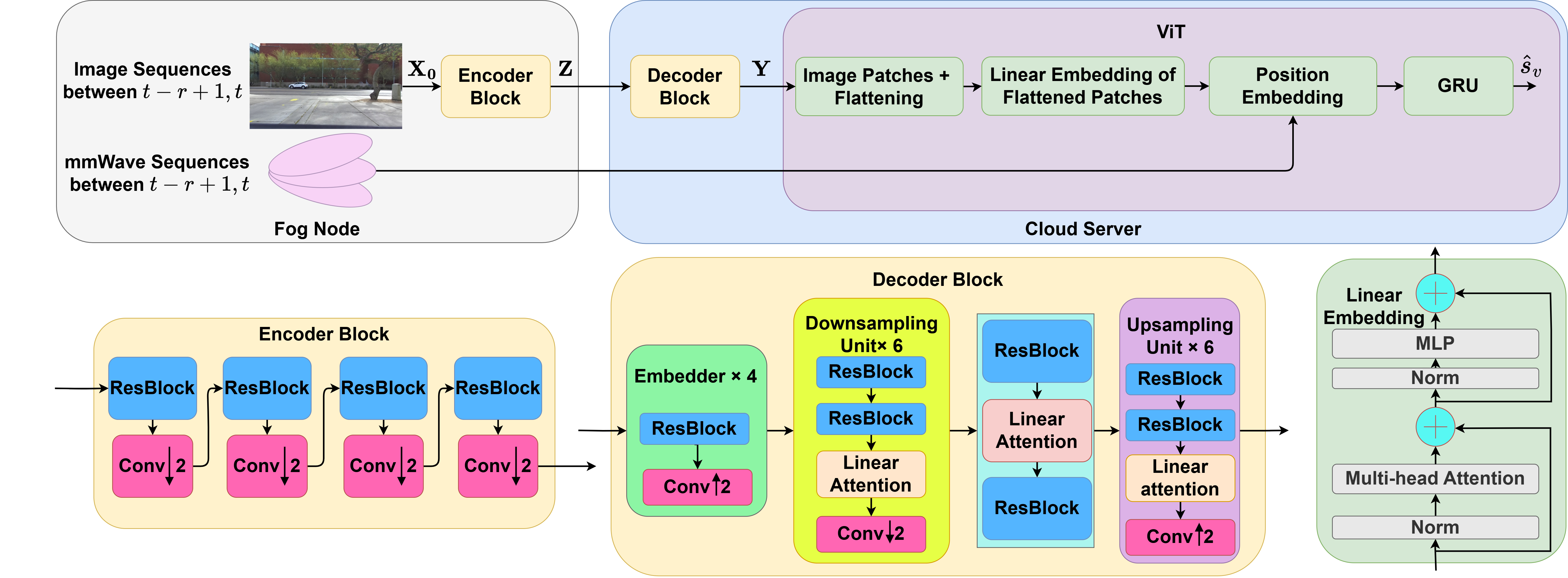}
  \captionof{figure}{Visual-aided blockage prediction framework at the fog-cloud system, processed through an encoder-decoder architecture and ViT.}
  \label{fig:image2}
\end{figure*}

In this section, we present the detailed implementation of our proposed method, as depicted in \figurename{\ref{fig:image2}}. The proposed method focuses on predicting blockages between the BS and the user by utilizing vision-aided data, including reconstructed RGB images and received mWave power vector information. In the following sections, we present the techniques for image compression and blockage prediction using ViT.

\subsection{Image Compression}
We perform image compression based on the conditional diffusion model \cite{yang2024lossy} to reduce the overhead of the fog-cloud system. The whole process includes two stages: the encoder stage and the decoder and reconstruction stage. We explain the implementation details of each stage as follows:

\textit{1) Encoder and Prior} 
The model employs a semantic latent variable $\mathbf{Z}$ to capture core image content and a set of texture variables $\mathbf{X}_{1:N}$ that encode the residual information, where $N$ denotes the number of decoding steps. For simplicity, we omit $t$ in this context and assume $\mathbf{X}_{0} = \mathbf{X}$. The joint probability distribution that describes the relationship between these variables is expressed as:
\begin{equation}
p(\mathbf{X}_{0:N}, \mathbf{Z}) = p(\mathbf{Y} | \mathbf{Z}) p(\mathbf{Z}),
\end{equation}
where $p(\mathbf{Y} | \mathbf{Z})$ represents the likelihood of reconstructing the original image given the semantic content $\mathbf{Z}$, and $p(\mathbf{Z})$ serves as the prior distribution for $\mathbf{Z}$.

The neural encoder $e(\mathbf{Z}|\mathbf{X})$ encodes the input image into a latent space and generates the semantic variable $\mathbf{Z}$. To enable efficient storage and transmission of $\mathbf{Z}$, a hierarchical prior distribution $p(\mathbf{Z})$ is utilized. This prior aids in enhancing entropy coding efficiency after $\mathbf{Z}$ is quantized and ensures that the compression process remains effective.

The primary objective of the encoder is to optimize the rate-distortion (R-D) trade-off, which involves finding a balance between the data compression rate and the quality of the reconstructed image. This balance is controlled through a Lagrange multiplier, denoted by $\lambda$, which adjusts the trade-off to achieve the desired performance.
The optimization process of the model is carried out by minimizing the negative modified Evidence Lower Bound, expressed as \cite{yang2024lossy}:
\begin{equation}
\resizebox{.87\hsize}{!}{$\mathcal{L}(\lambda, \mathbf{X}) = \mathcal{Q} + \lambda\mathcal{R} = \mathbb{E}_{\mathbf{Z}\sim e(\mathbf{Z}|\mathbf{X})}[-\log p(\mathbf{X}|\mathbf{Z}) - \lambda \log p(\mathbf{Z})]$},
\end{equation}
where $\mathcal{Q}$ denotes the distortion term, which quantifies the difference between the original and the reconstructed images, while $\mathcal{R}$ represents the bitrate term, which represents the cost associated with encoding the latent variable $\mathbf{Z}$. The expectation $\mathbb{E}_{\mathbf{Z}\sim e(\mathbf{Z}|\mathbf{X})}$ indicates that the average is taken over the distribution of $\mathbf{Z}$ as generated by the encoder.

Once the training phase is complete, the data encoding employs deterministic components, represented as $
\hat{\mathbf{Z}} = \lfloor\mathbf{Z}\rceil = \lfloor\text{Enc}(\mathbf{X})\rceil$,
where $\hat{\mathbf{Z}}$ denotes the rounded integer vector corresponding to the latent variable $\mathbf{Z}$.
To convert the continuous distribution $p(\mathbf{Z})$ into a discrete form, we use the transformation of $
P(\hat{\mathbf{Z}}) = \text{CDF}_p(\hat{\mathbf{Z}} + 0.5) - \text{CDF}_p(\hat{\mathbf{Z}} - 0.5),$
where $\text{CDF}_p$ represents the cumulative distribution function of $p(\mathbf{Z})$ \cite{yang2024lossy}. 

\textit{2) Decoding and Reconstruction Process} 
The decoder, following a conditional denoising diffusion model, performs the task of reconstructing $\mathbf{Y}$ based on the latent variable $\mathbf{Z}$. By applying Jensen's inequality, we can establish an upper bound for the R-D objective as follows \cite{yang2024lossy}:
\begin{equation}
\begin{split}
\mathbb{E}_{\mathbf{Z}\sim e(\mathbf{Z}|\mathbf{Y})}[-\log p(\mathbf{Y}|\mathbf{Z}) - \lambda \log p(\mathbf{Z})] \\ 
\leq \mathbb{E}_{\mathbf{Z}\sim e(\mathbf{Z}|\mathbf{Y})}[L_{\text{upper}}(\mathbf{Y}|\mathbf{Z}) - \lambda \log p(\mathbf{Z})],
\end{split}
\end{equation}
where $L_{\text{upper}}(\mathbf{Y}|\mathbf{Z})$ quantifies the model's capacity to reconstruct the image based on $\mathbf{Z}$. In contrast, $-\log p(\mathbf{Z})$ represents the number of bits required to encode $\mathbf{Z}$ according to its prior distribution. We simplify $L_{\text{upper}}(\mathbf{Y}|\mathbf{Z})$ by using the denoising score matching loss as follows:
\begin{equation}
\resizebox{.87\hsize}{!}{$L_{\text{upper}}(\mathbf{Y}|\mathbf{Z}) \approx \mathbb{E}_{\mathbf{Y},n,\epsilon} \left[\frac{\alpha_{n}}{1 - \alpha_{n}} \left\| \mathbf{Y} - \mathcal{X}_{\Phi} \left(\mathbf{X}_{n}, \mathbf{Z}, \frac{n}{N_{\text{train}}}\right) \right\|^{2} \right]$},
\end{equation}
where $n \sim \text{Unif}\{1,..., N\}$, $\epsilon \sim \mathcal{N}(0, \mathbf{I})$, $\mathbf{x}_n(\mathbf{x}_0) = \sqrt{\alpha_n}\mathbf{x}_0 + \sqrt{1 - \alpha_n}\epsilon$, $\alpha_n = \prod_{i=1}^n (1 - \beta_i)$, $\beta_n \in (0, 1)$, and $N_{\text{train}}$ is the total number of training steps. The function $\mathcal{X}_{\Phi}$ learns to reconstruct the original image directly from the noisy version. 

In the cloud server, the latent variable $\mathbf{Z}$ is decoded using prior $p(\mathbf{Z})$ and reconstructs image $\mathbf{Y}$ through ancestral sampling with stochastic noise $\mathbf{X}_N \sim \mathcal{N}(0, \gamma^2 \mathbf{I})$. Since the texture variables $\mathbf{X}_{1:N}$ are generated, varying the noise level parameter $\gamma$ results in different image reconstructions.
\subsection{Blockage Prediction using ViT}
After decoding, ViT extracts features from the images, which are then processed by a time-series classification component for binary decision-making. A Gated Recurrent Unit (GRU) network is utilized to capture long-term dependencies.

\subsection{DeepSense 6G Dataset}
We evaluate our proposed method using Scenario 21 of the DeepSense 6G dataset. The BS is a dual-band transmitter equipped with sensors that capture environmental information at a rate of 6.5 samples per second. These configurations simulate real-world conditions for beam management and blockage prediction, offering a comprehensive framework for developing and testing machine learning models.

\section{Simulation and Results}
\subsection{Parameter Settings}
To create $\mathcal{D}$ described in Section \ref{subsectionSystemModel}, we consider $r^{'} = r = 5$ and generate the time series dataset consisting of 5 input image samples and the corresponding future image samples. This means that our prediction method can consider within 750 ms to determine whether a blockage will be detected. We maintain a balanced dataset, ensuring a similar number of LOS and NLOS data sequences. The dataset is divided into training, validation, and test sets, following a distribution of 70\%, 20\%, and 10\%, respectively.

For image reconstruction, the noise level parameter $\gamma$ was set to 0.5 for all simulations, ensuring consistent reconstruction behavior across different testing conditions.

We directly use the pre-trained parameters for ViT provided by \cite{dosovitskiy2020image}  and then fine-tune them with the DeepSense 6G dataset. Each input image is divided into non-overlapping patches of 16x16. These patches are then linearly embedded in a 512-dimensional representation. Received mmWave power vector information is then added to these features to help the network identify the spatial location of each patch within the image. The results are fed into the transformer encoder, consisting of self-attention layers and fully connected neural blocks. 
Self-attention enables the model to focus on various areas of the data while capturing global relationships between patches. The ViT model is fine-tuned with a learning rate of 1e-4 and a batch size of 16 for the real-world Deepsense 6G dataset. 

Our GRU model consists of a single layer with 128 hidden units, employing Tanh activation and a dropout rate of 0.5 to prevent overfitting. The final output is obtained through a fully connected layer with a single neuron, utilizing a sigmoid activation function for binary classification. We opted for the Adam optimizer to handle the training process and conducted training over 100 epochs.

As a baseline, we use a CNN for feature extraction and a transformer for blockage prediction, with 100 epochs, a 0.0005 learning rate, and a batch size of 6.

The BS and the user are placed $d = 10.6$ meters apart, facing opposite directions. The transmit signal power is set to $P_{t}$ = 40 dBm, and the background noise power is set to $N_{0}$ = -10 dBm. 
$h_k[t]$ is influenced by two key factors at $t$. The first factor is frequency $f$, which is impacted by the free-space path loss model expressed as $PL_{LOS} = 20 \log_{10}(d) + 20 \log_{10}(f) + 20 \log_{10}\left(\frac{4\pi}{c}\right)$, where $d$ denotes the distance between transmitter and receiver and $c$ denotes the speed of light. The second factor is blockage, which causes attenuation in $h_k[t]$ across different frequencies.
For mmWave, we use $f = 28$ GHz and bandwidth $BW = 5$ GHz and for Sub-6 GHz, we use $f = 2.5$ GHz and $BW = 500$ MHz
In mmWave frequencies, blockages cause additional attenuation, significantly affecting the signal-to-noise ratio (SNR) at the receiver. In contrast, Sub-6 GHz frequencies offer better propagation characteristics and lower attenuation due to their longer wavelength \cite{3GPPmmWave}. These characteristics make Sub-6 GHz particularly suitable for scenarios with blockages. 

\subsection{Simulation Results}
\begin{figure}\setlength{\belowcaptionskip}{-6pt}
  \begin{subfigure}{0.23\textwidth}
    \includegraphics[width=\linewidth]{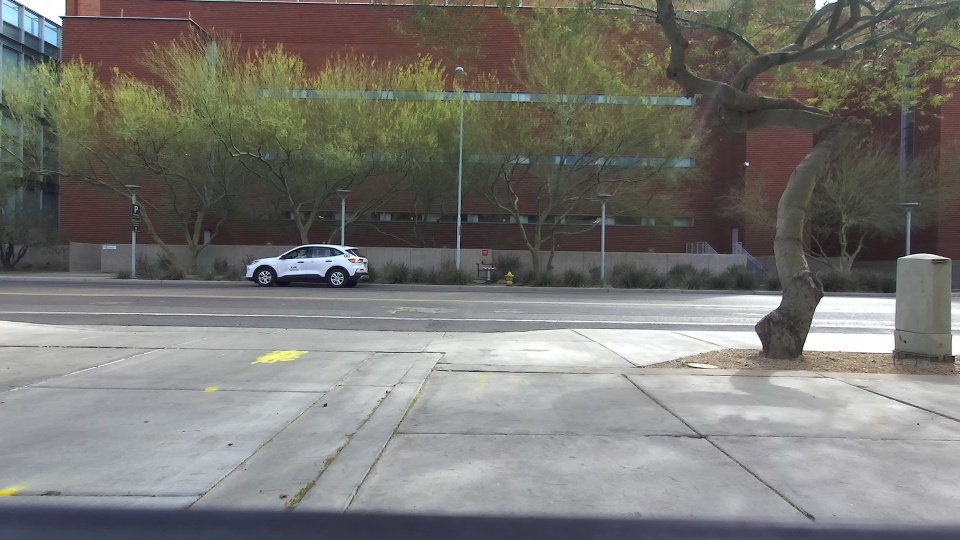}
    \caption{Ground Truth} 
  \end{subfigure}%
  \hspace*{\fill}   
  \begin{subfigure}{0.23\textwidth}
    \includegraphics[width=\linewidth]{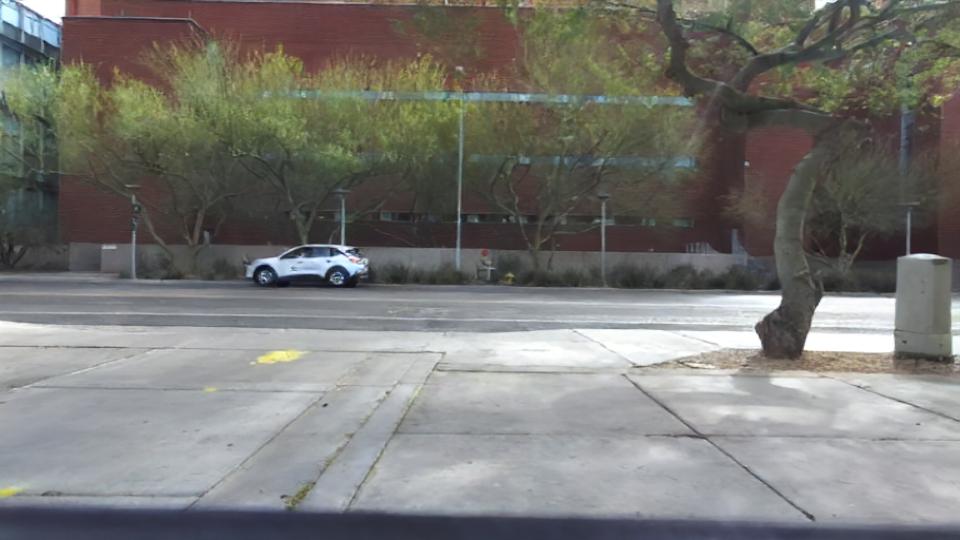}
    \caption{After Compression} 
  \end{subfigure}%
\caption{Image compression comparison.} \label{fig:IMAGE3BPP}
\end{figure}

\figurename{\ref{fig:IMAGE3BPP}} demonstrates the visual impact of our compression technique on a sample image. \figurename{\ref{fig:IMAGE3BPP}}.a shows the original uncompressed ground truth, while \figurename{\ref{fig:IMAGE3BPP}}.b displays the result after compression.
The original images were initially compressed using JPEG as a lossy format, and we further compressed them using the method introduced in \cite{yang2024lossy}.
The compression method achieves an average compression ratio of 0.2969 for the latent variable $\mathbf{z}$. This translates to a significant 70.31\% reduction in bandwidth requirements when transmitting information from fog nodes to cloud centers.
As evident in \figurename{\ref{fig:IMAGE3BPP}}.b, the decoded result exhibits a slight degradation in quality compared to the original. However, this minor degradation has minimal impact on blockage prediction accuracy.

Table \ref{tab:blockage_prediction_accuracy} compares the accuracy of different blockage prediction methods. 
According to the results, the ViT-based approach achieved the highest accuracy of 94.23\%. Although integrating compressed images slightly reduced the accuracy to 92.78\%, it significantly improved data efficiency. Similarly, transformers had an accuracy of 88.12\%, while using compressed images led to a small reduction to 86.23\%. These results demonstrate that the use of image compression can improve data efficiency with only a minimal impact on prediction accuracy.

\figurename{\ref{fig:IMAGE4Loss}} illustrates the loss values for each method, where the ViT approach exhibited the lowest loss, supporting its superior accuracy. In contrast, Transformers with compressed images had the highest loss, indicating reduced prediction performance. Overall, although image compression enhances the efficiency of offloading tasks, it results in a slight reduction in blockage prediction accuracy.
\begin{table}[ht]\setlength{\belowcaptionskip}{-6pt}
    \centering
    \caption{Blockage Prediction Accuracy}
    \begin{tabular}{|c|c|}
        \hline
        \textbf{Blockage Prediction Method} & \textbf{Accuracy} \\
        \hline
        Transformers - no image compression & 88.12\% \\
        \hline
        Transformers with image compression & 86.23\% \\
        \hline
        ViT  - no image compression & 94.23\% \\
        \hline
        Proposed Method & 92.78\% \\
        \hline
    \end{tabular}
    \label{tab:blockage_prediction_accuracy}
\end{table}

\begin{figure}
\setlength{\belowcaptionskip}{-12pt}
\centering
  \centering
  \includegraphics[width=0.85\linewidth]{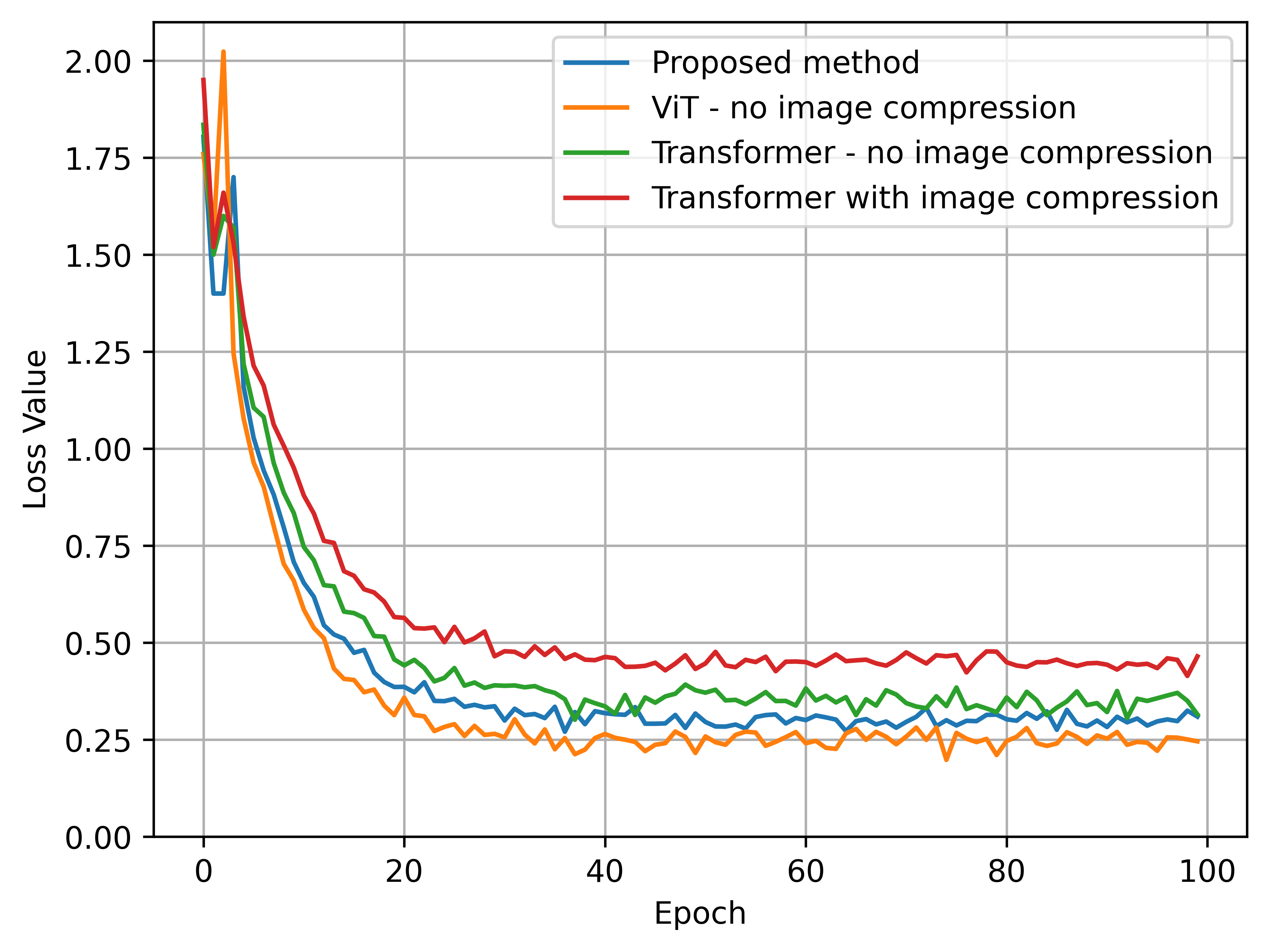}
  \captionof{figure}{Loss value comparison of blockage prediction methods.}
  \label{fig:IMAGE4Loss}
\end{figure}

\begin{figure}
\setlength{\belowcaptionskip}{-12pt}
\centering
  \centering
  \includegraphics[width=0.85\linewidth]{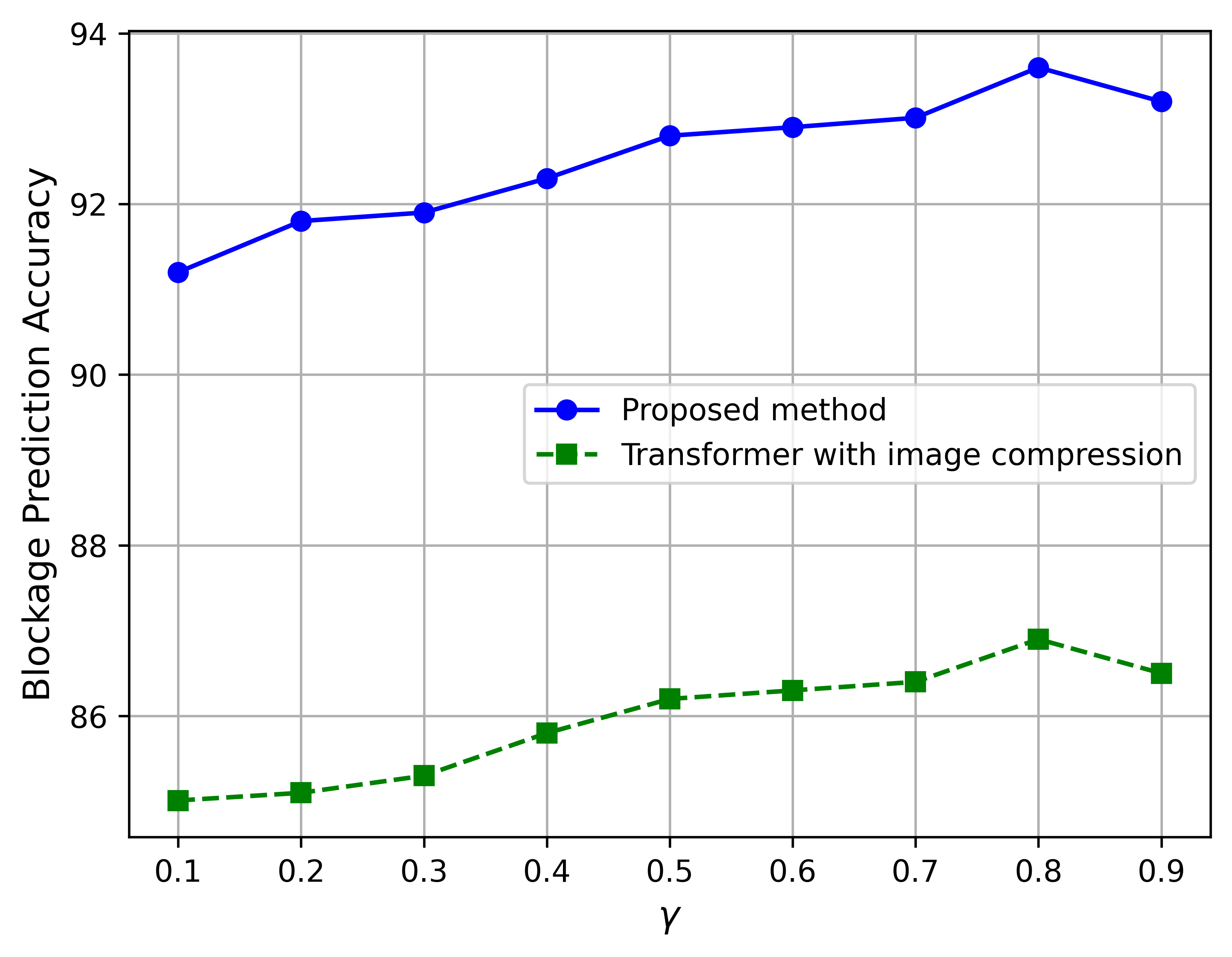}
  \captionsetup{belowskip=-12pt}
  \captionof{figure}{Effect of $\gamma$ on blockage prediction accuracy.}
  \label{fig:IMAGE5NoiseImpact}
\end{figure}

By adjusting the noise level parameter $\gamma$ during the image decoding process, we can achieve different decoding outcomes.
The effect of $\gamma$ on the accuracy of blockage prediction is shown in \figurename{\ref{fig:IMAGE5NoiseImpact}}, where $\gamma = 0.5$ used in previous simulations resulted in lower prediction accuracy. Our findings indicate that values of $\gamma$ greater than 0.8 lead to improved perceptual quality and higher distortion. In terms of finding the optimal balance, a $\gamma$ value of 0.8 offers the best blockage prediction accuracy outcomes for our proposed method.

\begin{figure}
\setlength{\belowcaptionskip}{-12pt}
\centering
  \centering
  \includegraphics[width=0.85\linewidth]{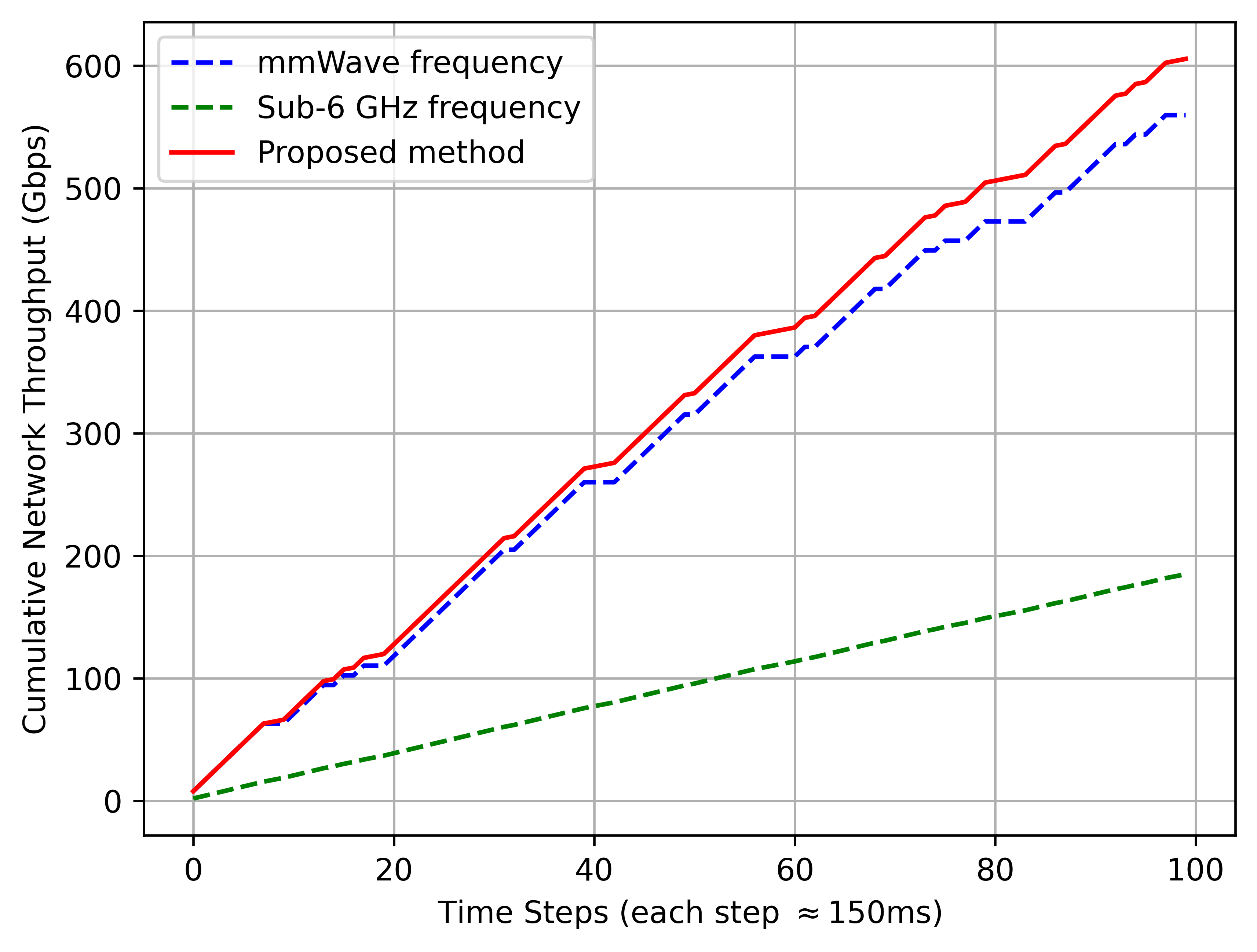}
  \captionsetup{belowskip=-12pt}
  \captionof{figure}{Comparison of cumulative network throughput over time.}
  \label{fig:IMAGE6CumulativeNetworkthroughput}
\end{figure}

\figurename{\ref{fig:IMAGE6CumulativeNetworkthroughput}} compares the cumulative network throughput across different time steps. We calculate throughput according to Shannon capacity $C = BW \cdot \log_{2}(1 + SNR)$. We assumed that 30 blockages occurred randomly across 100 steps.
As observed from \figurename{\ref{fig:IMAGE6CumulativeNetworkthroughput}}, the mmWave frequency system offers high data rates when there are no blockages, but its performance drops significantly during blockages. In contrast, the Sub-6 GHz system maintains more stable performance, though it has lower overall data rates. This is because the mmWave system is highly susceptible to blockages, which cause severe signal degradation, whereas the Sub-6 GHz frequencies have better penetration and can sustain connections even in non-line-of-sight conditions. For instance, at step 40, a blockage caused the cumulative network throughput to remain unchanged, indicating no data transmission between the transmitter and receiver. Sub-6 GHz frequency system, however, is more stable but with lower data rates. The proposed method switches between mmWave and Sub-6 GHz frequencies based on the blockage prediction accuracy. 
The results indicate that the proposed method leverages both technologies to achieve a slight improvement in cumulative network throughput.

\begin{figure}
\setlength{\belowcaptionskip}{-12pt}
\centering
  \centering
  \includegraphics[width=0.85 \linewidth]{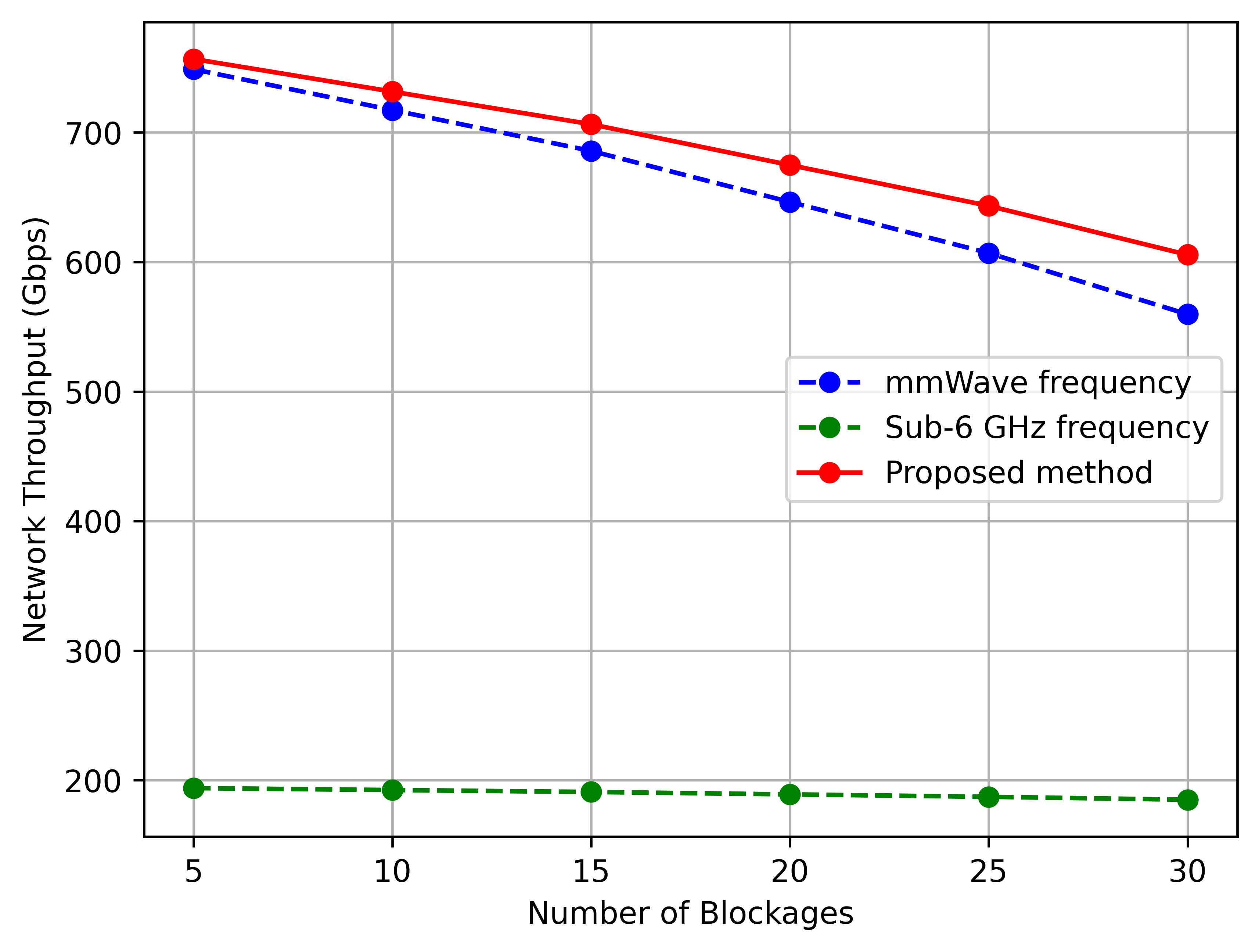}
  \captionsetup{belowskip=-12pt}
  \captionof{figure}{Comparison of cumulative network throughput under varying number of blockages.}
  \label{fig:IMAGE7CumulativeNetworkthroughputVSNumber}
\end{figure}

\figurename{\ref{fig:IMAGE7CumulativeNetworkthroughputVSNumber}} illustrates the impact of varying the number of blockages, ranging from 5 to 30, on network throughput. As expected, the network throughput decreases as the number of blockages increases. However, the proposed method demonstrates slightly better performance, especially in environments with a high density of blockages.

\begin{figure}
\setlength{\belowcaptionskip}{-12pt}
\centering
  \centering
  \includegraphics[width=0.85\linewidth]{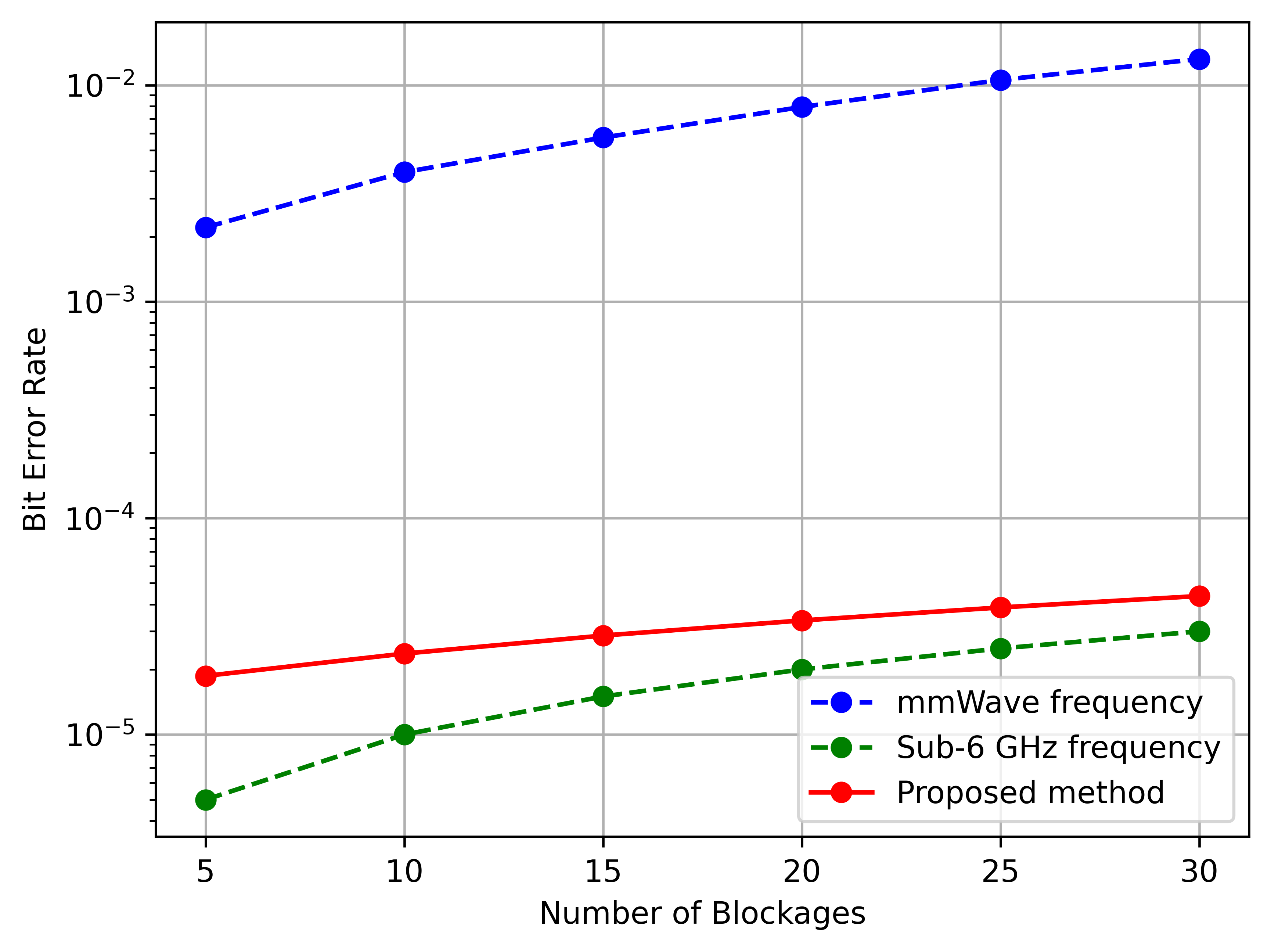}
  \captionsetup{belowskip=-12pt}
  \captionof{figure}{BER Comparison under varying numbers of blockages}
  \label{fig:IMAGE8BER}
\end{figure}

\figurename{\ref{fig:IMAGE8BER}} presents the bit error rate (BER) under different blockage conditions. In our analysis, we assume the use of quadrature phase shift keying (QPSK) modulation, with the bit error rate (BER) defined as $\text{BER} = \frac{1}{2} \text{erfc} \left( \sqrt{\frac{\text{SNR}}{2}} \right)$, where $\text{erfc}$ denotes the complementary error function. The results show that the mmWave frequency system exhibits poor performance as the number of blockages increases, while the Sub-6 GHz frequency system maintains a robust connection. Importantly, the proposed method maintains a BER close to that of the Sub-6 GHz frequency system, emphasizing its resilience even under challenging conditions.

\section{Conclusion}
In this paper, we proposed a visual-aided framework for blockage prediction and dual-band communication. Our proposed method integrates generative AI-based image compression within a hierarchical fog-cloud architecture, addressing both the computational challenges and bandwidth constraints inherent in visual data. 
By leveraging ViT, our method achieves a blockage prediction accuracy of 92.78\% on the real-world DeepSense 6G dataset, outperforming the baseline transformer model's 86.23\% accuracy with compressed images, while also reducing transport bandwidth by 70.31\%. This approach improves network throughput and reliability, particularly in environments with high blockage density

\section*{Acknowledgment}
This work has been supported by NSERC Canada Research Chairs program, MITACS, and Ericsson.

\bibliographystyle{IEEEtran}
\bibliography{bibliography.bib}

\end{document}